\documentclass[slac_one]{revtex4}
\usepackage{graphicx}
\usepackage[abs]{overpic} 
\usepackage{fancyhdr}
\def \rightdownarrow
 {\kern.3em
 \rule[.5ex]{.15mm}{2ex}
 {\mbox{$\kern-0.1em{\longrightarrow}$}}      
 }

\def\lessim{\mathrel {\vcenter {\baselineskip 0pt \kern 0pt  
\hbox{$<$} \kern 0pt \hbox{$\sim$} }}}

\def\gessim{\mathrel {\vcenter {\baselineskip 0pt \kern 0pt   
\hbox{$>$} \kern 0pt \hbox{$\sim$} }}}

\newcommand{\ie}{i.~e.,}					

\newcommand{\pipi}{\ensuremath{\pi\pi}}

\newcommand{\mpipi}{\ensuremath{m_{\pipi}}}

\newcommand{\Lumi}{\ensuremath{\mathcal{L}}}			

\newcommand{\lumifb}{\mbox{fb$^{-1}$}}				

\newcommand{\tev}{\ensuremath{\mathrm{Te\kern -0.1em V}}}
\newcommand{\gev}{\ensuremath{\mathrm{Ge\kern -0.1em V}}}	
\newcommand{\mev}{\ensuremath{\mathrm{Me\kern -0.1em V}}}	
\newcommand{\kev}{\ensuremath{\mathrm{ke\kern -0.1em V}}}	
\newcommand{\massgev}{\mbox{\gev/$c^2$}}			
\newcommand{\massmev}{\mbox{\mev/$c^2$}}			
\newcommand{\pgev}{\mbox{\gev/$c$}}				

\newcommand{\stat}{\ensuremath{\mathit{~(stat.)}}}		
\newcommand{\syst}{\ensuremath{\mathit{~(syst.)}}}		

\newcommand{\CP}{{\rm CP}}                                            

\newcommand{\bd}{\ensuremath{B^{0}}}				
\newcommand{\bs}{\ensuremath{B^{0}_s}}				

\newcommand{\bu}{\ensuremath{B^{+}}}				

\newcommand{\bhadron}{\mbox{$b$-hadron}}			

\newcommand{\bn}{\ensuremath{B^{0}_{(s)}}}			

\newcommand{\fullbhh}{\ensuremath{B \to hh'}}

\newcommand{\Fig}[1]{Figure \ref{fig:#1}}

\newcommand{\fig}[1]{fig.~\ref{fig:#1}}

\newcommand{\dedx}{\ensuremath{\mathit{dE/dx}}}

\newcommand{\ptot}{\ensuremath{p_{\rm{tot}}}}

\newcommand{\etal}{et al.}

\def\babar{\mbox{\slshape B\kern-0.1em{\smaller A}\kern-0.1em B\kern-0.1em{\smaller A\kern-0.2em R}}}

\newcommand{\Bhh}{\ensuremath{\bn \rightarrow h^{+}h^{'-}}}

\newcommand{\BdKpi}{\ensuremath{\bd \rightarrow K^{+} \pi^-}}

\newcommand{\BsKpi}{\ensuremath{\bs \rightarrow K^- \pi^+}}

\newcommand{\Bspipi}{\ensuremath{\bs \rightarrow  \pi^+ \pi^-}}

\newcommand{\BdKK}{\ensuremath{\bd \rightarrow  K^+ K^-}}

\newcommand{\BR}{\ensuremath{\mathcal B}}

\def\beq{\begin{equation}}
\def\eeq{\end{equation}}
\def\bea{\begin{eqnarray}}
\def\eea{\end{eqnarray}}

\def\sss{\scriptscriptstyle}

\def\barp{{\raise.35ex\hbox
{${\sss (}$}}---{\raise.35ex\hbox{${\sss )}$}}}
\def\bdbarp{\hbox{$B_d$\kern-1.4em\raise1.4ex\hbox{\barp}}}
\def\bsbarp{\hbox{$B_s$\kern-1.4em\raise1.4ex\hbox{\barp}}}

\def\roughly#1{\mathrel{\raise.3ex\hbox
{$#1$\kern-.75em\lower1ex\hbox{$\sim$}}}}

\newcommand{\Bd}{\ensuremath{B^{0}}}

\newcommand{\Bs}{\ensuremath{B_{s}^{0}}}
\newcommand{\aBs}{\ensuremath{\overline{B}_{s}^{0}}}

\newcommand{\Lb}{\ensuremath{\Lambda_{b}^{0}}}

\newcommand{\bear}{\begin{array}}
\newcommand{\ear}{\end{array}}
\newcommand{\bet}{\begin{tabular}}
\newcommand{\eet}{\end{tabular}}
\newcommand{\beqn}{\begin{eqnarray}}
\newcommand{\eeqn}{\end{eqnarray}}

\newcommand{\cdf}{CDF Collaboration}

\begin{document}

\title{Hadronic $B_{s}$ decays} 

%

\author{F. Ruffini, on behalf of the CDF Collaboration}
\affiliation{INFN of Pisa and University of Siena, Italy\\
Polo Fibonacci, Largo B. Pontecorvo, 56127 Pisa - Italy\\
E-mail: ruffini@pi.infn.it
}

\begin{abstract}
The study of b-hadrons made possible a great number of benchmark results in flavour physics. In recent years
a lot of effort in understanding their dynamics has been done. These studies can represent a possible avenue for the 
discovery of physics beyond the Standard Model (SM) in less studied sectors, and can as well 
be used as a tool to properly test the hadronic calculations reliability. In this sense, a unique opportunity 
is represented by the $B_{s}$ mesons. They are less studied and known with respect to 
$B^{+}$ and $B^{0}$ mesons, but high precision measurement are possible with the current 
available statistics at b-factories and at hadronic machines. 
Here we present a brief collection of recent $B_{s}$ mesons results.
\end{abstract}

\maketitle

\thispagestyle{fancy}


\section{$B \to h^{+}h^{-'}$ DECAYS}
Non-leptonic two-body charmless decays of neutral $B$ mesons (\Bhh, where $h$ is a charged pion or kaon)
are very interesting for the understanding
of flavor physics and CP violation mechanism in the B-meson sector. 
Their rich phenomenology offered several opportunities to measure and constrain
the parameters of the quark-mixing matrix (\ie\ Cabibbo-Kobayashi-Maskawa, CKM).
These processes also allow to access the phase of the $V_{ub}$ element of the 
CKM matrix ($\gamma$ angle), and to test the reliability of the SM and hadronic calculations.
The presence of New Physics (NP) can be revealed by its impact on their decay amplitudes, where new particles may enter in penguin diagrams.

The large production cross section of $b$ hadrons of all kinds at the TeVatron and LHC
allows extending our knowledge of the \bs\ and \Lb\ decays,
which are important to supplement our understanding of \bd\  and \bu\ meson decays provided by the $B$-factories.
The branching fraction of \BsKpi\ decay mode provides information on the CKM angle 
$\gamma$~\cite{Gronau:2000md} and the measurement of direct
CP asymmetry could be a powerful model-independent test 
of the source of CP asymmetry in the $B$ system \cite{Lipkin:2005pb}.
The \Bspipi\ and \BdKK\  decay modes proceed through annihilation and exchange topologies, which
are currently poorly known and a source of significant uncertainty in many theoretical calculations~\cite{B-N,Bspipi}. 
A measurement of both decay modes would allow a determination of the strength of these amplitudes~\cite{Burasetal}.
Indeed, the CDF Collaborations reported the first evidence for \Bspipi\ decay mode and set a limit on the Branching Ratio \BR(\BdKK) \cite{CDF_bspipi_6fb}.
The LHCb Collaboration confirmed the evidence and also obtained the observation for the \Bspipi\ and set a limit on \BR(\BdKK) \cite{:2012as}. 

The measurements performed at the two hadron-machines face the same challenge, 
\ie\ to disentangle different decay modes overlapping into a single peak.
In this conditions, every mode is a background for the other signals. The analysis strategies are similar: the idea is to disentangle the singular components using kinematic and Particle Identification (PID) information. While at CDF the PID information allows a limited statistical separation and it is necessary combining kinematics and PID information using a 5-dimensional Maximum Likelihood Fit to obtain the results, LHCb benefits from the different detector structure and skills. In particular, the presence of the RICH detectors allows a powerful particle identification. The efficiency in identifying the final states particles is good enough to make possible measurements competitive (and better) with the CDF ones performing only a fit on the mass distribution.

\subsection{CDF analysys}
CDF analyzed an integrated luminosity  
$\int\Lumi dt\simeq 6$~\lumifb\ sample of pairs of oppositely-charged particles
with $p_{T} > 2$~\pgev\ and $p_{T}(1) + p_{T}(2) > 5.5$~\pgev, used to form $B$ candidates.
The trigger required also a transverse opening angle $20^\circ < \Delta\phi < 135^\circ$ between the two tracks, to reject background from particle pairs within the same jet and from back-to-back jets.
In addition, both charged particles were required to originate from
a displaced vertex with a large impact parameter (100 $\mu$m $< d_0(1,2) < 1$~mm), 
while the \bhadron\ candidate was required to be produced in
the primary $\bar{p}p$ interaction ($d_0< 140$~$\mu$m) and to have travelled a transverse distance
$\L_{T}>200$~$\mu$m. A sample of about 3 million 
 \fullbhh\ decay modes  
(where  $ B = \Bd,\Bs ~{\rm or}~ \Lb$  and $h= K~{\rm or}~ \pi$) 
was reconstructed after the off-line confirmation of trigger requirements. 
In the offline analysis, an unbiased optimization procedure determined a
tightened selection on track-pairs fit to a common decay vertex.
%
The offline selection is based on a more accurate 
determination of the same quantities used in the trigger.

No more than one $B$ candidate per event is found after this 
selection, and a mass ($m_{\pi\pi}$) is assigned to 
each, using a charged pion mass assignment for both decay products. 

\begin{figure}[htb]
\centering
\begin{overpic}[scale=0.35]{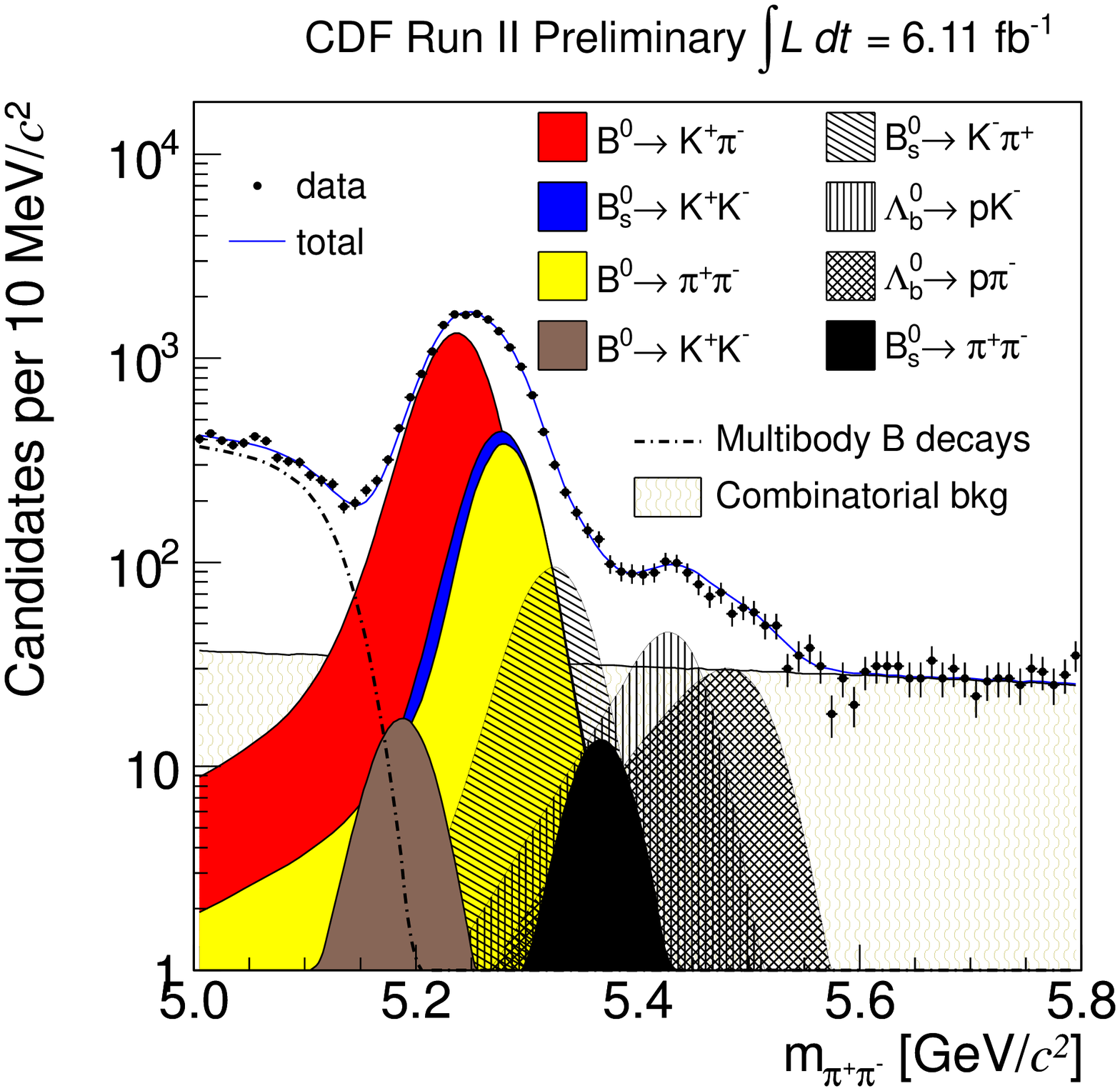}
\put(5,180){(a)}
\end{overpic}   
\begin{overpic}[scale=0.35]{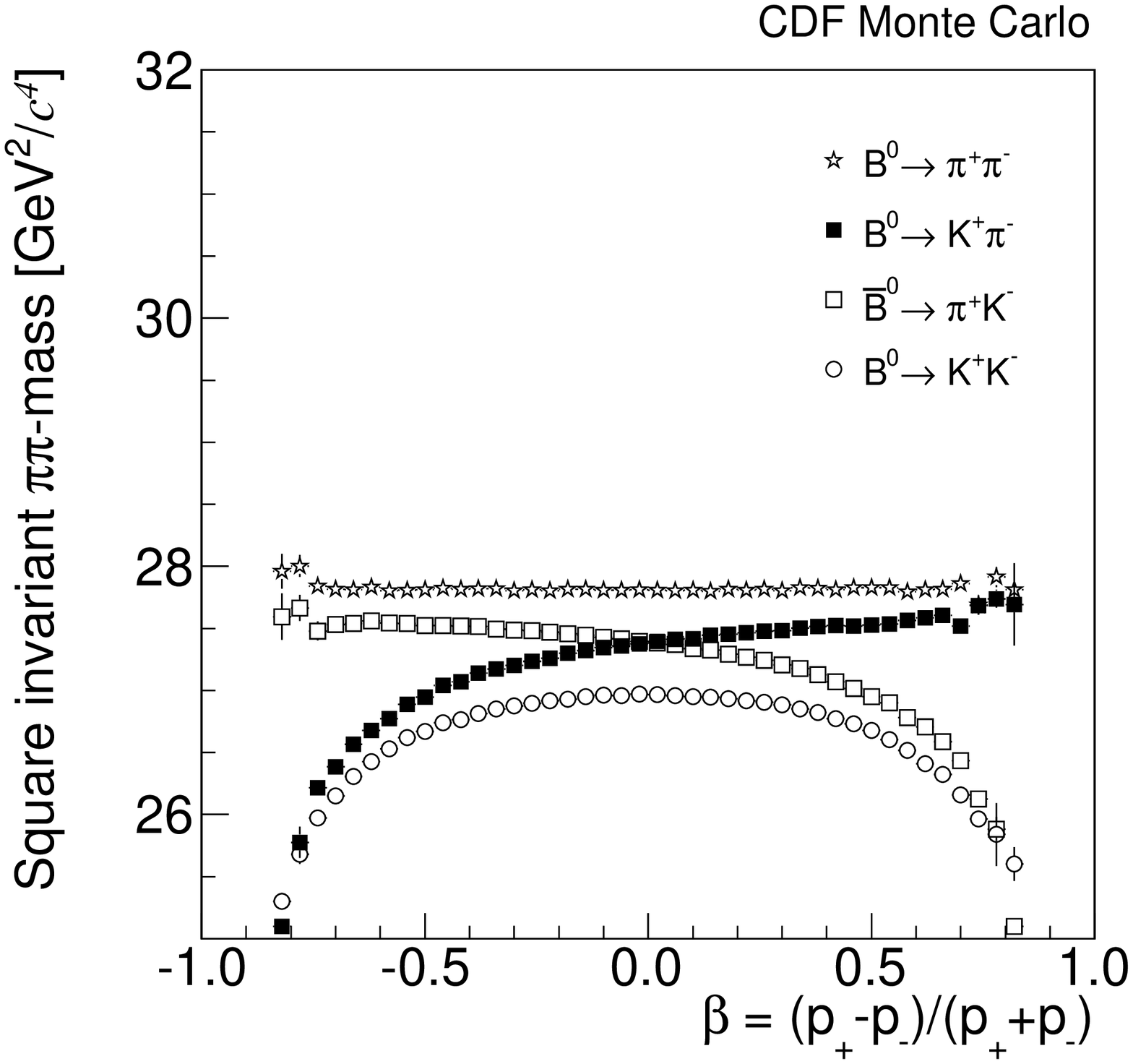}
\put(5,180){(b)}
\end{overpic}   
\caption{Invariant mass distribution of \fullbhh\ candidates passing cuts selection,  using a pion mass assumption for both decay products. Cumulative projections of the Likelihood fit for each mode are overlaid (a). Profile plots of the square invariant $\pi\pi$-mass as a function of charged momentum asymmetry $\beta$ for all \Bd\ simulated signal modes (b).} 
\label{fig:projections}
\end{figure}

The resulting $\pi\pi$-mass distributions (see Figure~\ref{fig:projections}, (a)) show a clean signal of \fullbhh\ decays.
Backgrounds include mis-reconstructed multibody $b$-hadron decays (physics background, causing the enhancement at \mpipi $< 5.16$ \massgev) and random pairs of charged particles (combinatorial background).
In spite of a good mass resolution ($\approx 22\,\massmev$), the various \fullbhh\ modes overlap into an unresolved
mass peak near the nominal \Bd\ mass, with a width of about $\approx 35\,\massmev$. 

The resolution in invariant mass and in particle identification, provided by specific ionization energy loss  (\dedx) in the drift chamber, is not sufficient for separating the individual \fullbhh\ decay modes on an event-by-event basis,
therefore a Maximum Likelihood fit, incorporating kinematics and PID information, was performed.  The kinematic information is summarized by three loosely correlated
observables: (a) the square of the invariant $\pi\pi$ mass
$m^{2}_{\pi\pi} $; (b) the signed momentum imbalance
$\beta = (p_{+} - p_-)/(p_+ + p_{-})$, where $p_+$ ($p_-$) is the
the momentum of the positive (negative) particle; (c) the scalar sum of particle momenta 
$\ptot=p_+ + p_-$.

The likelihood exploits the kinematic differences among modes (see Figure~\ref{fig:projections}, (b)) by using the correlation between the signed momenta of the tracks and the invariant masses $m^{2}_{+-}$ of a candidate for any mass assignment of the decay products ($m_{+}$,$m_{-}$), using the equation
\begin{eqnarray}\label{eq:Mpipi2}
 m^{2}_{+-}  =  m^{2}_{\pi\pi}  -  2 m_{\pi}^2 + m_{+}^2+m_{-}^2 +
                 \nonumber    \\
          -  2 \sqrt{p_{+}^2+m_{\pi}^2} \sqrt{p_{-}^2+m_{\pi}^2}  
                    +  2\sqrt{p_{+}^2+m_{+}^2} \sqrt{p_{-}^2+m_{-}^2}.
 \end{eqnarray} 
This procedure is useful to obtain statistical separation power between $\pi\pi$ and $KK$ (or $K\pi$) final states, and therefore is a key tool for the measurement of the observables of interest.
Kinematic fit templates are extracted from simulation for signal and physics background, while they are extracted from an independent sample
for combinatorial background.
Mass line-shapes are accurately described according for the non Gaussian resolution tails and for the effects  of the final state radiation of the soft photons. The \dedx\ is calibrated over the tracking volume and time using about 3.2 millions of $D^{*+} \to D^{0} [\to K^{-}\pi^{+}]\pi^{+}$ decays, where the sign of the soft pion tags the $D^{0}$ flavor. A 1.4$\sigma$ separation is obtained between kaons and pions with $p > 2$ \pgev, becoming 2.0$\sigma$ for the couples $KK/\pi\pi$ and $K^{+}\pi^{-}/K^{-}\pi^{+}$. 
 \dedx\ templates (signal and background) are extracted from the $D^{0}$ samples used in calibration.


The dominant contributions to the systematic uncertainty are the uncertainty on the \dedx\ calibration and parameterization and 
the uncertainty on the combinatorial background model. An additional systematic uncertainty, of the order of 10\% has been assessed because of a fit bias, found in the estimate of the relative fraction of the \BdKK\ decay mode. 
Other contributions come from trigger efficiencies, physics background shape and kinematics, $b$--hadron masses and lifetimes.

The signal yields are calculated from the signal fractions returned by the likelihood fit.
For the first time significant signal is seen for \Bspipi, with a significance of $3.7\sigma$, while the significance for the \BdKK\ decay mode is $2.0$ $\sigma$. 


Absolute results are listed in Table~\ref{tab:summary_Bspipi}; they are obtained 
by normalizing the data to the world--average of \BR(\BdKpi)~\cite{pdg_2010}. 
A 90\% of confidence level interval is also quoted for the \BdKK\ mode. 

The branching fraction of the \BdKK\ mode is in agreement with other existing measurements~\cite{pdg_2010},  while it is higher than the predictions~\cite{Beneke:2003zv}\cite{Cheng:2009cn}.

The branching fraction of the \Bspipi\ mode is in agreement with the theoretical expectations within the pQCD approach \cite{Ali:2007ff}, \cite{Li:2004ep} while is higher than most other predictions ~\cite{Beneke:2003zv} \cite{Sun:2002rn} \cite{Chiang:2008vc} \cite{Cheng:2009mu}.

\subsection{LHCb analysis}
We report preliminary measurements on by LHCb collaboration  using an integrated luminosity of 0.32 \lumifb\ collected during the first part of 2011 at centre of mass energy of 7 TeV. The \Bhh\ decays events used are extracted from the triggered data using different offline selections, each one targeted to achieve the best sensitivity on the measurements of interest.
The key point of the analysis is the PID information: the PID observables allows to discriminate between the various decay modes. Hence, in order to determine 
the amount of cross-feed backgrounds for a given channel, the relative efficiencies of the PID 
selection cuts, employed to identify the specific final state of interest, play a key role. 
Huge statistics sample of $D^{*+}\to D^{0}(K\pi)\pi^{+}$ +c.c., decay modes with similar kinematic features of the \Bhh, are used to extract the PID efficiency. Since the mass region greater then 5.6 \massgev\ are characterized by the presence of $\Lambda^{0}_{b}\to p \pi(K)$ decay modes, the PID information has also been studied for protons particles using a sample of $\Lambda^{0}\to p\pi^{-}$ decays.

\begin{figure}[t]
\begin{center}
\includegraphics[width=0.36\textwidth]{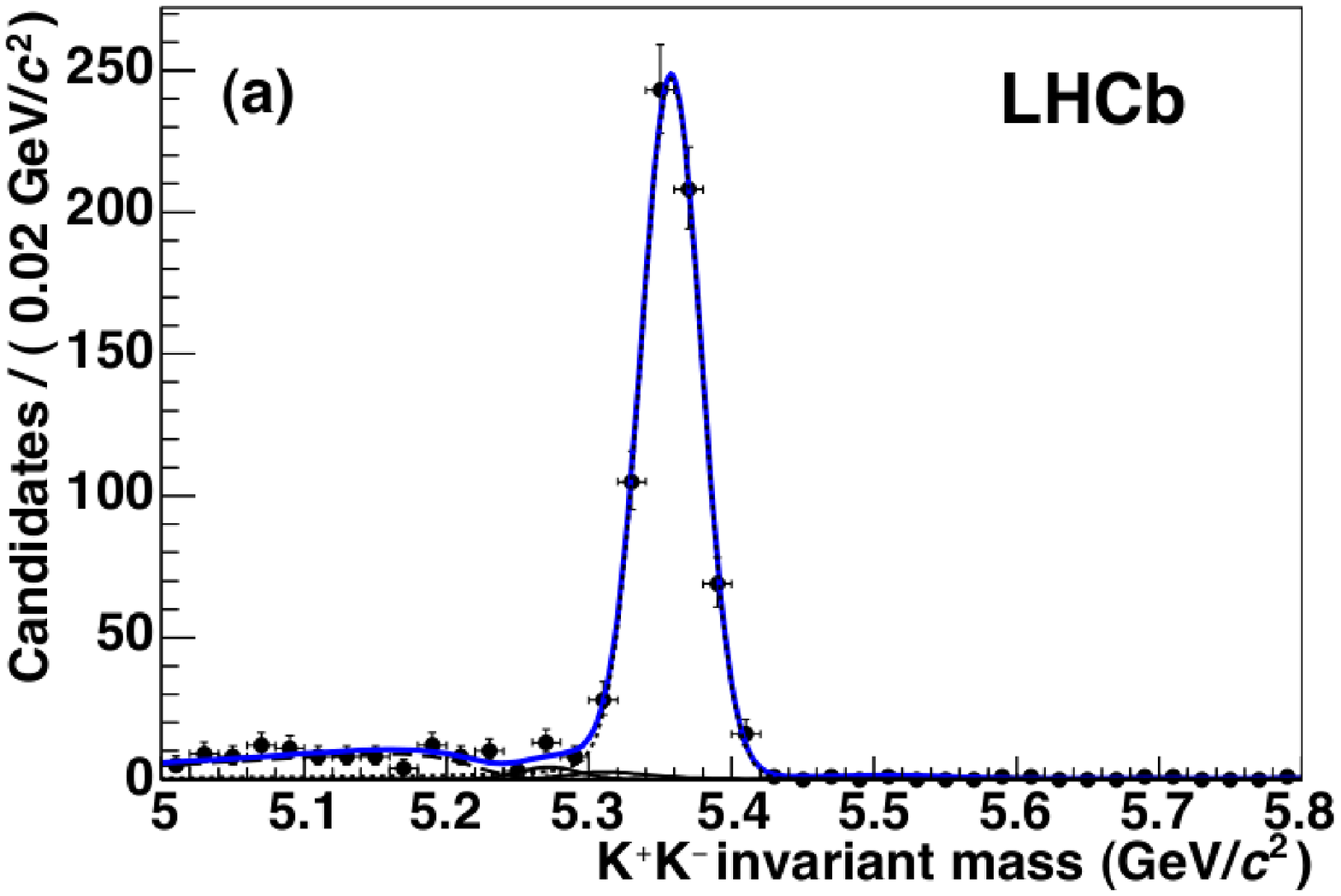} \includegraphics[width=0.36\textwidth]{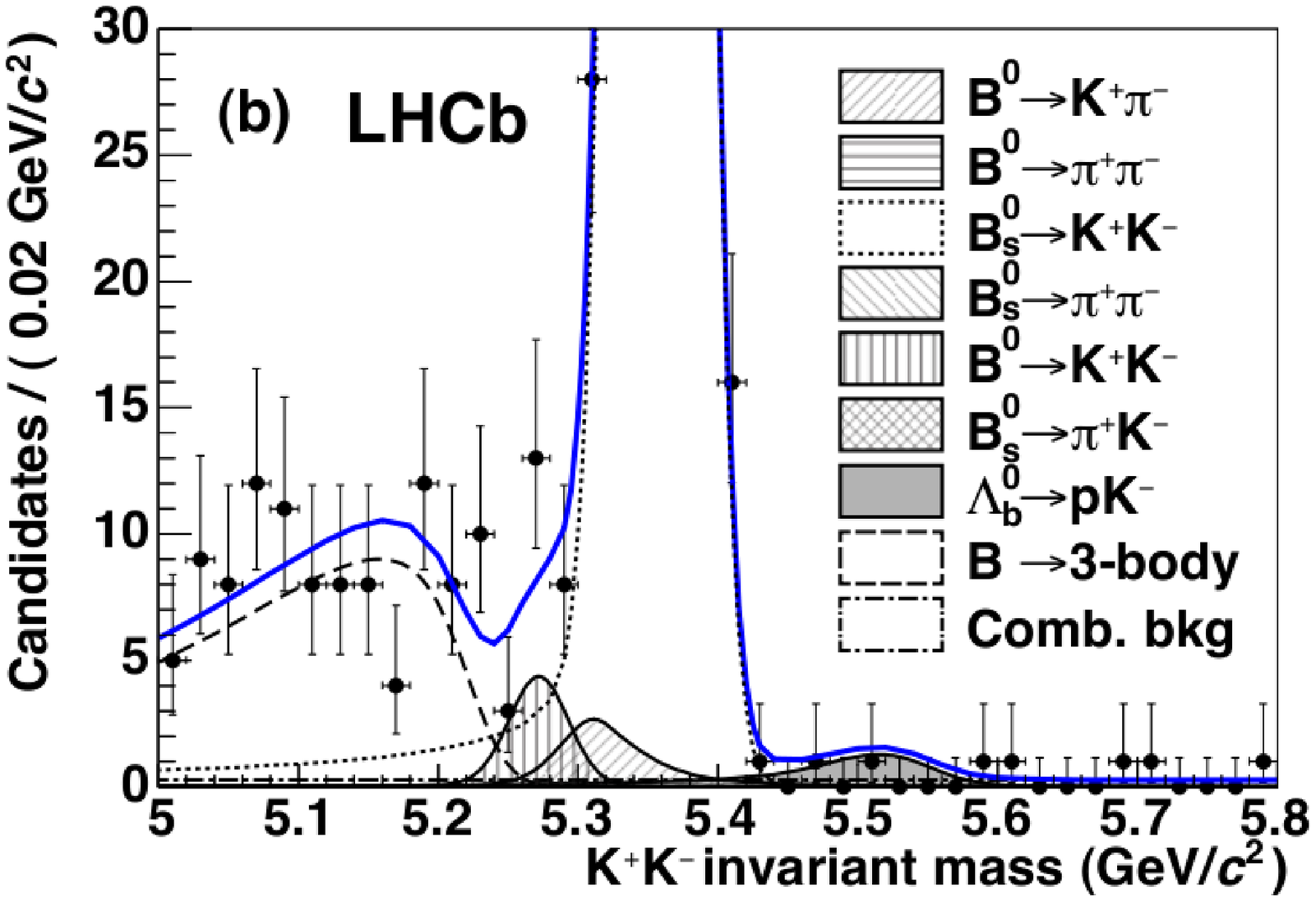}
\includegraphics[width=0.36\textwidth]{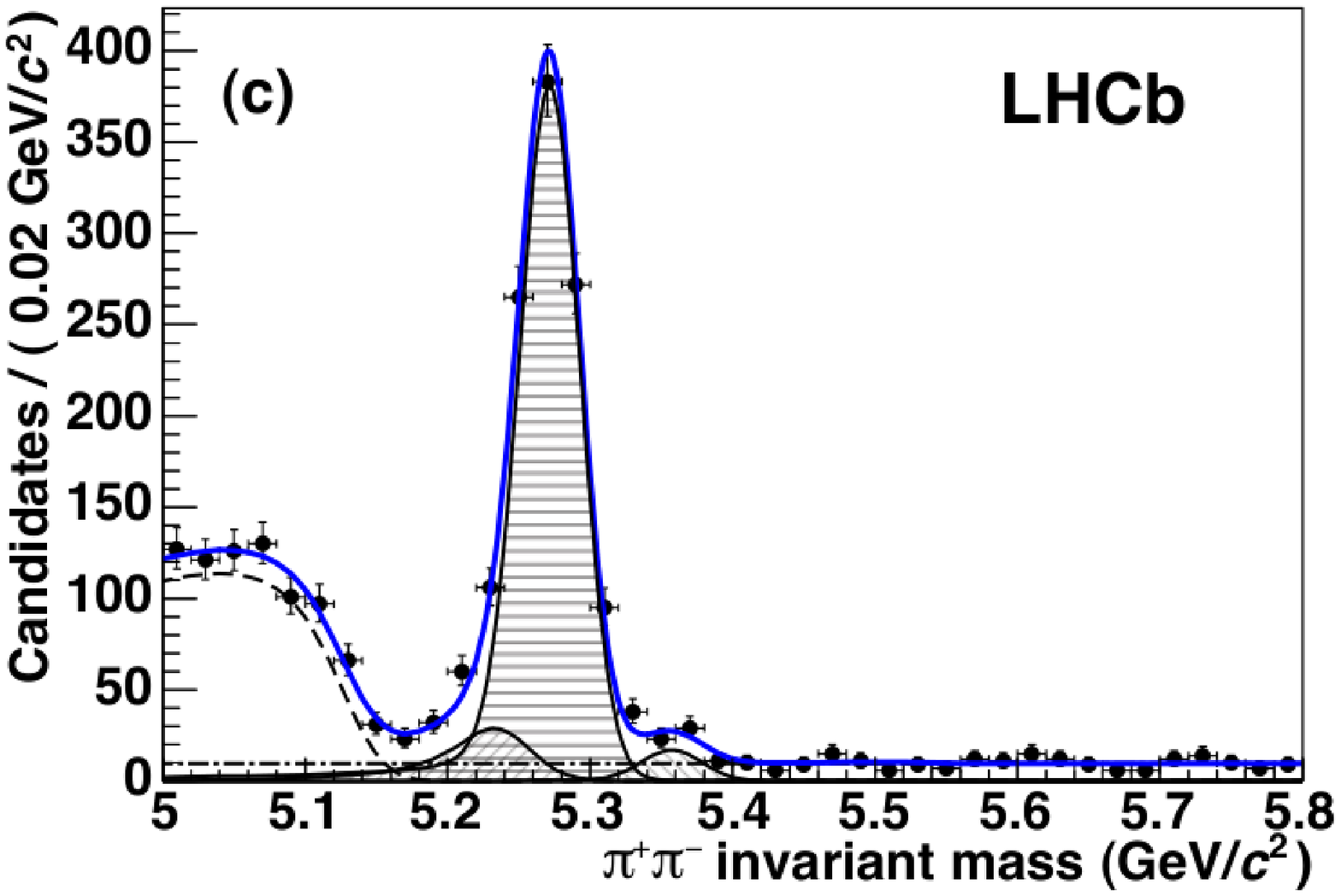} \includegraphics[width=0.36\textwidth]{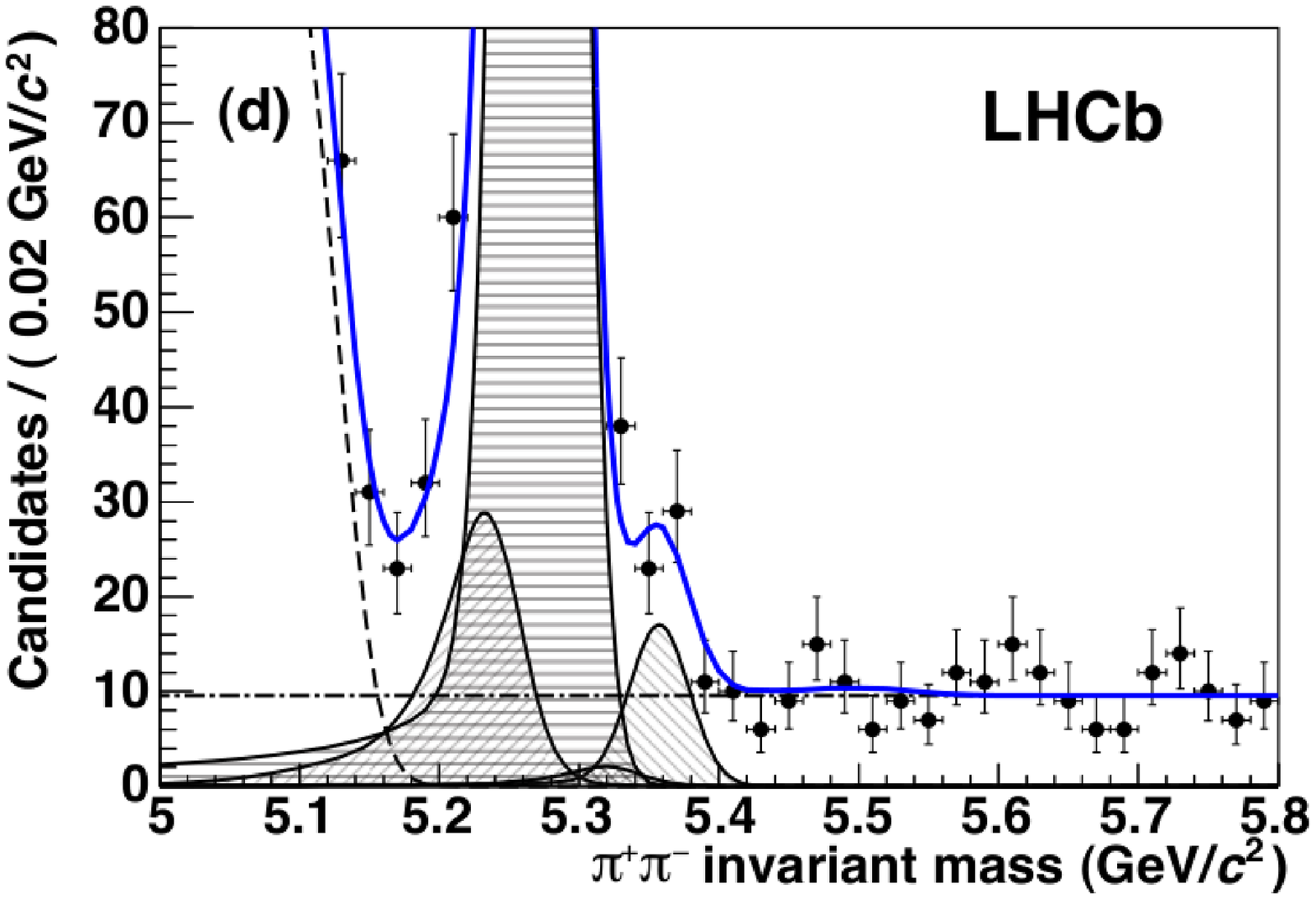}
\end{center}
\caption{Invariant mass distributions for the mass hypotheses (a, b) $K^+K^-$ and (c, d) $\pi^+\pi^-$. Plots (b) and (d) are the same as (a) and (c) respectively, but magnified to focus on the rare $B^0 \rightarrow K^+ K^-$ and $B^0_s \rightarrow \pi^+ \pi^-$ signals. The results of the unbinned maximum likelihood fits are overlaid. The main components contributing to the fit model are also shown.}
\label{fig:raredecays}
\end{figure}

The \BdKK\ and \Bspipi\ signal events are extracted with an unbinned 
maximum likelihood fit to different final state mass spectra of events passing the offline selections. 
The various final states are separated using PID requirements in order to have exclusive categories 
corresponding to distinct final state hypotheses ( $K^{+}\pi^{-}, K^{-}\pi^{+}, K^{+}K^{-}$ or $\pi^{+}\pi^{-}$). 
In the fits the amount of background for a given channel, due to the other channels where at least one 
particle has been mis-identified (cross-feed background), has been taken into account. The result 
of the fit is shown on \Fig{raredecays}.

In particular, LHCb find a significance of more than 5$\sigma$ for the \Bspipi\ signal, representing the first observation of this decay mode. LHCb results are in agreement with CDF results, with about the same level of precision. 

%
%
%
%

\begin{table}[h!]
{
{
\scriptsize
\begin{tabular}{l|cc}
\hline
\hline
Mode          &  Absolute \BR (10$^{-6}$)  	&  Absolute \BR (10$^{-6}$) \\
	          &  CDF  					& LHCb\\
\hline
\BdKK        & 0.23 $\pm$ 0.10 $\pm$ 0.10   & 0.13 $^{+0.06}_{-0.05}$ $\pm$ 0.07 \\
\Bspipi        & 0.57 $\pm$ 0.15 $\pm$ 0.10   & 0.98 $^{+0.23}_{-0.19}$ $\pm$ 0.11 \\
\hline
\hline
\end{tabular}
}
}
\caption{\label{tab:summary_Bspipi} Absolute branching fractions results for CDF \cite{CDF_bspipi_6fb} and LHCb \cite{:2012as}.}
\end{table}

\section{$\Bs \to J/\psi f_{0}$(980) lifetime}
In the standard model, the mass and flavor eigenstates of the \Bs\ meson differ. This gives rise to particle-antiparticle oscillations, which proceed in the SM through weak interaction processes, and whose phenomenology depends on the Cabibbo-Kobayashi-Maskawa (CKM) quark mixing matrix. The time evolution of \Bs\ mesons is governed by a Schr\"odinger equation which contains two 2x2 matrices, called mass and decay matrix. The off-diagonal elements are related to observable quantities, namely the mass difference $\Delta m_s$, the phase $\phi_s$ between the off-diagonal elements of mass and decay matrices, and the decay width difference $\Delta \Gamma_s$, depending on the decay matrix elements and on the phase $\phi_s$.  A special feature of the \Bs\ system is the large value of $\Delta \Gamma_s$, which yields a significant difference in the lifetimes of the two mass eigenstates of the \Bs. Within the standard model $\phi_s$ is predicted to be very small, which in consequence means that CP and mass-eigenstates coincide. 
If new physics is present, it could enhance $\phi_s$ to large values, a 
scenario which is not excluded by current experimental 
constraints. In such a case the correspondence between 
mass and CP eigenstates does not hold anymore and the 
measured lifetime will correspond to the weighted average of the lifetimes 
of the two mass eigenstates with 
weights dependent on the size of the CP violating phase 
$\phi_s$. Thus a measurement of the \Bs\ lifetime in a final 
state which is a CP eigenstate provides, in combination 
with other measurements, valuable information on the 
decay width difference $\Delta \Gamma_s$ and the CP violation in \Bs\ mixing. 

Using $p\bar{p}$ collision data with an integrated luminosity of 3.8 \lumifb\ collected by the CDF II detector at the Tevatron it is possible to measure the $\Bs$ lifetime using \Bs\ decays to the CP-odd final state $J/\psi f_{0}$(980) with $J/\psi \to \mu^{+}\mu^{-}$ and $f_{0}$(980) $\to \pi^{+}\pi^{-}$ \cite{Aaltonen:2011nk}.

\begin{figure}[htb]
\centering
\includegraphics[scale=0.35]{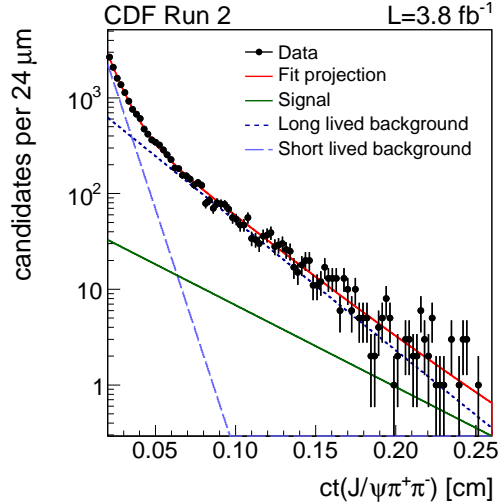}
\caption{Decay time distribution with fit projection overlaid. }
\label{fig:projections_Bslifetime}
\end{figure}

To extract the \Bs\ lifetime CDF used a maximum Likelihood fit on three variables:
the invariant mass, the decay time and the decay time uncertainty of each 
candidate. An accurate study of the templates of the signals and of the combinatorial background
was performed. \Fig{projections_Bslifetime} reports the projection on the life time variable. 
The dominant contributions to the systematic uncertainty are the uncertainties related to the background models, for the mass template and for the decay time template.

The result obtained from the fit is: 
\begin{equation}
\tau(\Bs \to J/\psi f_{0}(980)) = 1.70 ^{-0.11}_{+0.12} \stat \pm 0.03 \syst \mbox{ps}. 
\end{equation}
This is the first measurement of the \Bs\ lifetime in a decay to a CP eigenstate and corresponds in the standard model to the lifetime of the heavy \Bs\ eigenstate. 
 The measured value agrees well 
both with the standard model expectation as well as with 
other experimental determinations. 
While the precision of the lifetime measurement is 
still limited by statistics, it provides an important cross-check on the result determined in 
$\Bs \to J/\psi \phi$ decays,
which relies on an angular separation of two CP eigenstates. Furthermore, the measured lifetime can be used 
as an external constraint in the $\Bs \to J/\psi \phi$ 
analysis  to improve the determination of the CP-violating phase. The lifetime measurement 
in $\Bs \to J/\psi f_0$(980) decays is also the next step towards a tagged time dependent CP violation measurement, which can provide an independent constraint on 
the CP violation in the mixing.

\section{$\Bs \to D_{s}^{(*)+}D_{s}^{(*)-}$ DECAYS}
A \Bs\ meson can oscillate into its antiparticle via second order weak interaction transitions, which make its time 
evolution sensitive to contributions from new physics processes. Such contributions are not well constrained yet 
and might be responsible for the deviation from the standard model reported in \cite{Abazov:2011yk}. The \Bs\ eigenstates with 
defined mass and lifetime, $B^0_{sL}$ and $B^0_{sH}$ are linear combinations of the \Bs\ and  \aBs\ states and, in the standard model, correspond in good approximation to the even and odd CP eigenstates, respectively. In the absence of substantial CP violation, a sizable decay width difference between the light and heavy mass eigenstates, 
$\Delta \Gamma_s = \Gamma_{sL} - \Gamma_{sH}$, arises from 
the fact that decays to final states of definite CP are only accessible by one of the mass eigenstates. The dominant 
contribution to $\Delta \Gamma_s$ is believed to come from the
$\Bs \to D_{s}^{(*)+}D_{s}^{(*)-}$ (c.c. modes implicitly included), which are predominantly CP even 
and saturate $\Delta \Gamma_s$  under certain theoretical assumptions \cite{Aleksan:1993qp,Shifman:1987rj}, resulting in the relation 
\begin{equation}
2\BR( \Bs \to D_{s}^{(*)+}D_{s}^{(*)-} ) \approx \frac{\Delta \Gamma_s}{ \Gamma_s + \Delta \Gamma_s /2}
\label{2BR_gamma}
\end{equation}
where $\Gamma_s = (\Gamma_{sL} + \Gamma_{sH})/2$  \cite{Dunietz:2000cr} . However, three-body modes may provide a significant contribution to $\Delta \Gamma_s$  \cite{Chua:2011er}. 
A finite value of $\Delta \Gamma_s$  improves the experimental sensitivity to CP violation because it allows one to distinguish the 
two mass eigenstates via their decay time distribution. Furthermore, the $\Bs \to D_{s}^{(*)+}D_{s}^{(*)-}$
 decays could be used in 
future to measure directly the lifetime of the CP -even eigenstate, which would complement the CP odd eigenstate 
lifetime measurement in $\Bs \to J/\psi f^0$ (980) as we mentioned before and provide additional information in the search for new physics contributions to CP violation.
The $\Bs \to D_{s}^{(*)+}D_{s}^{(*)-}$ decays have been previously studied by the ALEPH, CDF, D0, and Belle collaborations \cite{Barate:2000kd, Abulencia:2007zz, Abazov:2008ig, Esen:2010jq}. Here we present the latest preliminary result from Belle and from CDF. 
\subsection{Belle results}
At the Belle, the decays $\Bs \to D_{s}^{(*)+}D_{s}^{(*)-}$ are reconstructed in a 
data sample corresponding to an integrated luminosity of 121.4 \lumifb\ at the $\Upsilon$(5s) resonance. 
The reconstruction of $D^+$ candidates has been made using six final states: $\phi \pi^+$, $K^0K^+$, $ K^{*0} K^+,$ $\phi \rho^+$, $K^0 
K^{*+},$ and $K^{*0} K^{*+}$. 
Belle combined the $D^{+}$ candidates with photon 
candidates to reconstruct $D^{*+}_s \to D^+_s \gamma$ decays. The strategy of the analysis involve the use of a two dimensional Likelihood fit to separate the signal from background. The variables of the fit are the beam-energy-constrained mass $M_{bc} = \sqrt{E^2_{\mbox{beam}} - p^2_B}$, and the energy difference 
$\Delta E = E_B - E_{\mbox{beam}}$, where $p_B$ and $E_B$ are the reconstructed momentum and energy of the \Bs\ candidate, and $E_{\mbox{beam}}$ is the beam energy. All signal template are extracted from the simulations and calibrated using $\Bs \to D_s^{(*)-}\pi^+$ and $\Bs \to D_s^{(*)+}D^-$ decays. 

The results \cite{Esen:2011in} are $\Bs \to D_{s}^{+}D_{s}^{-} = (0.6 \pm0.1 \pm 0.1)\%$,   $\Bs \to D_{s}^{*\pm}D_{s}^{*\mp} = (1.8 \pm 0.2 \pm 0.4)\%$ and  $\Bs \to D_{s}^{*+}D_{s}^{*-} = (2.0 \pm0.3 \pm 0.5)\%$, the latest corresponding to the first observation.

 Assuming these decay modes saturate decays to CP even final states, the branching fraction determines the relative width difference 
between the \Bs\ \CP\ odd and even eigenstates. Taking \CP\ violation to be negligibly small, we obtain 
\begin{equation}
\Delta \Gamma_s /  \Gamma_s = (9.0\pm0.9\pm 2.2)\% 
\end{equation}
where $\Gamma_s$ is the mean decay width. 
This result is in good agreement with the current experimental measurements \cite{pdg_2010} and it is consistent with theory \cite{Lenz}.

\subsection{CDF results}
\begin{figure}[htb]
\centering
\includegraphics[width=0.24\textwidth]{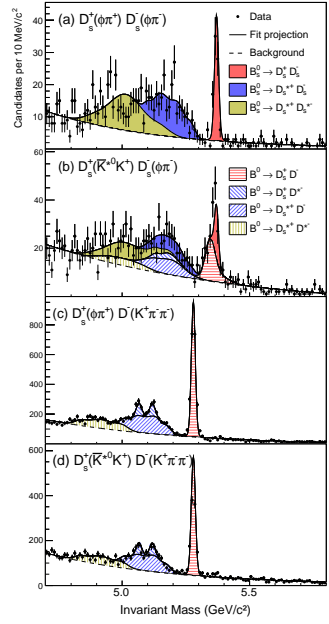}
\caption{Invariant mass distribution of
(a) $\Bs \rightarrow D^+_s(\phi \pi^+)  D^-_s(\phi \pi^-)$,
(b) $\Bs \rightarrow D^+_s(K^{*0} K^+)  D^-_s(\phi \pi^-)$,
(c) $\Bd \rightarrow D^+_s(\phi \pi^+)  D^-(K^+ \pi^- \pi^+)$, and
(d) $\Bd \rightarrow D^+_s(K^{*0} K^+)  D^-(K^+ \pi^- \pi^+)$ candidates
with the simultaneous fit projection overlaid.
The broader structures stem from decays where the photon or $\pi^0$ from the $D_{(s)}^{*+}$ decay is not reconstructed.
Misreconstructed signal events in (c) show up as reflections in (b).}
\label{fig:BsDsDs_plots}
\end{figure}

At CDF, the decays $\Bs \to D_{s}^{(*)+}D_{s}^{(*)-}$ with $D^{+}_{s} \to K^{-}K^{+}\pi^{+}$ are reconstructed in a data sample corresponding to an integrated luminosity of $\int\Lumi dt\simeq 6.8$~\lumifb\ sample collected by the detector.
CDF measured the $B_{s}$ production rate times the $\Bs \to D_{s}^{(*)+}D_{s}^{(*)+}$ branching ratio relative to the normalization mode $\Bd \to D_{s}^{+}D^{-}$. The strategy analysis is simple: the relative branching fractions are determined by a simultaneous fit on the mass variable to two signal and two normalization samples (\fig{BsDsDs_plots}). The values returned by the fit must be corrected for the relative efficiencies. An accurate determination of the efficiencies is achieved by taking into account the Dalitz structure of the $D^{+}_{s}$ decay. As a result CDF obtained \cite{Aaltonen:2012mg}: \BR($\Bs \to D_{s}^{+}D_{s}^{-}$) = ($0.49 \pm 0.06 \pm 0.05 \pm 0.08$)\%, \BR($\Bs \to D_{s}^{*\pm}D_{s}^{*\mp}$) = ($1.13 \pm 0.12 \pm 0.09 \pm 0.19$)\%, \BR($\Bs \to D_{s}^{*+}D_{s}^{*-}$) = ($1.75 \pm 0.19 \pm 0.27 \pm 0.29$)\% and \BR($\Bs \to D_{s}^{(*)+}D_{s}^{(*)-}$) = ($3.38 \pm 0.25 \pm 0.30 \pm 0.56$)\%. Statistical, systematic and normalization uncertainties are reported.
These measurements represents the world best measurements up to date.

The information on the branching ratios are useful to infer indirect information about $\Delta \Gamma_{s}$, using relation \ref{2BR_gamma}:
\begin{equation}
\Delta \Gamma_{s}/\Gamma_{s} = (6.99 \pm 0.54 \pm 0.64 \pm 1.20)\%
\end{equation}
which is consistent with SM expectation \cite{Lenz}.


\section{CONCLUSIONS}
These measurements represent a brief collection of different attempts to enter a new era: with the current statistics from b-factories and hadron machines, it is possible to undergo the systematic and detailed exploration of the $B_{s}$ sector. Many new results are expected to come: Belle and CDF are squeezing up their full data sample, while the advent of LHCb will make possible unprecedented statistical precision measurements.


\bibliographystyle{ieeetr}

\begin{thebibliography}{99} 
\bibliographystyle{ieeetr}
\bibitem{Gronau:2000md} M~Gronau and J.~L.~Rosner, 
Phys.\ Lett.\ {\bf B482}, 71 (2000), hep-ph/0003119.

\bibitem{Lipkin:2005pb} ~H.~J.~Lipkin, 
Phys.\ Lett.\ {\bf B621}, 126 (2005), hep-ph/0503022.

\bibitem{B-N} M.~Beneke and M.~Neubert, Nucl.\ Phys.\ {\bf B675}, 333 (2003), hep-ph/0308039.

\bibitem{Bspipi}
  Y.~D.~H.~Yang \etal, 
  Eur.\ Phys.\ J., {\bf C44}, 243 (2005), hep-ph/0507326.

\bibitem{Burasetal} A.~J.~Buras \etal, 
  Nucl.\ Phys.\ {\bf B697},  133 (2004), hep-ph/0402112.

\bibitem{CDF_bspipi_6fb} T. Aaltonen \etal\ (\cdf), 
Phys.\ Rev.\ Lett.\ {\bf 108}, 211803 (2012), arxiv/1111.0485 [hep-ex].

\bibitem{:2012as}
  RAaij {\it et al.}  [LHCb Collaboration],
  arXiv:1206.2794 [hep-ex].

\bibitem{pdg_2010} K. Nakamura \etal\,
J. Phys. G  {\bf 37}, 075021(2010). 

\bibitem{Beneke:2003zv}
M. Beneke and M. Neubert, Nucl. Phys.
 {\bf B675}, 333, 2003.

\bibitem{Cheng:2009cn}
H.-Y. Cheng, C.-K. Chua, Phys. Rev.
 D. {\bf 80}, 114008, 2009.

\bibitem{Ali:2007ff}
A. Ali \etal, Phys. Rev. D, {\bf 76} 074018,
2007.

\bibitem{Li:2004ep}
Y. Li. \etal,
Phys. Rev. D {\bf 70}, 034009, 2004.

\bibitem{Sun:2002rn}
J.-F. Sun \etal,
Phys. Rev. D {\bf 68}, 054003, 2003.

\bibitem{Chiang:2008vc}
C.-W. Chiang, M. Gronau and J. L. Rosner, 
Phys. Lett. B {\bf 664}, 169, 2008.

\bibitem{Cheng:2009mu}
H.-Y. Cheng, C.-K. Chua, Phys. Rev.
 D. {\bf 80}, 114026, 2009.

\bibitem{Aaltonen:2011nk}
  T.~Aaltonen {\it et al.}  [CDF Collaboration],
  Phys.\ Rev.\ D {\bf 84} (2011) 052012  [arXiv:1106.3682 [hep-ex]].



\bibitem{Abazov:2011yk} 
  V.~M.~Abazov {\it et al.} (D0 Collaboration),
  Phys.\ Rev.\ D {\bf 84}, 052007 (2011).

\bibitem{Aleksan:1993qp}
  R.~Aleksan {\it et al.},
  Phys.\ Lett.\ B {\bf 316}, 567 (1993).
\bibitem{Shifman:1987rj}
  M.~A.~Shifman and M.~B.~Voloshin,
  Yad.\ Fiz.\ {\bf 47}, 801 (1988) 
  [Sov.\ J.\ Nucl.\ Phys.\ {\bf 47}, 511 (1988)].
\bibitem{Dunietz:2000cr} 
  I.~Dunietz, R.~Fleischer and U.~Nierste,
  Phys.\ Rev.\ D {\bf 63}, 114015 (2001).
  
\bibitem{Chua:2011er}
  C.~K.~Chua, W.~S.~Hou, and C.~H.~Shen,
  Phys.\ Rev.\  D {\bf 84}, 074037 (2011).

\bibitem{Barate:2000kd}
  R.~Barate {\it et al.} (ALEPH Collaboration),
  Phys.\ Lett.\ B {\bf 486}, 286 (2000).
  
\bibitem{Abulencia:2007zz}
  T.~Aaltonen {\it et al.} (CDF Collaboration),
  Phys.\ Rev.\ Lett.\  {\bf 100}, 021803 (2008).
  
\bibitem{Abazov:2008ig}
  V.~M.~Abazov {\it et al.} (D0 Collaboration),
  Phys.\ Rev.\ Lett.\  {\bf 102}, 091801 (2009).
  
\bibitem{Esen:2010jq}
  S.~Esen {\it et al.} (Belle Collaboration),
  Phys.\ Rev.\ Lett.\  {\bf 105}, 201802 (2010).


\bibitem{Lenz} A. Lenz and U. Nierste, J. High Energy Phys. 0706, 072 
(2007); arXiv:1102.4274 [hep-ph].


\bibitem{Esen:2011in}
  S.~Esen,
  arXiv:1110.2099 [hep-ex].

\bibitem{Aaltonen:2012mg}
  T.~Aaltonen {\it et al.}  [CDF Collaboration],
  Phys.\ Rev.\ Lett.\  {\bf 108} (2012) 201801
  [arXiv:1204.0536 [hep-ex]].


\end{thebibliography}


\end{document}